\documentclass[twocolumn,pra,aps,groupedaddress,nofootinbib]{revtex4-1}
\usepackage{amsmath}
\usepackage{amsfonts}
\usepackage{amssymb}
\usepackage{graphicx}
\usepackage{hyperref}
\usepackage{bm}
\usepackage{color}
\usepackage{times}
\hypersetup{backref,
colorlinks=true,
linkcolor=blue,
linktoc=black,
citecolor=cyan,
urlcolor=black}

\begin{document}
\title{
Adiabatic spin-dependent momentum transfer in an $SU(N)$ degenerate Fermi gas
}

\author{P. Bataille$^1$}
\author{A. Litvinov$^1$}
\author{I. Manai$^1$}
\altaffiliation{Now at Laboratoire de Physique des Lasers, Atomes et Mol\'ecules (PhLAM), CNRS, Universit\'e de Lille.}
\author{J. Huckans$^2$}
\author{F. Wiotte$^1$}
\author{A. Kaladjian$^1$}
\author{O. Gorceix$^1$}
\author{E. Mar\'echal$^1$}
\altaffiliation{Now at Laboratoire de Physique et d'Etude des Mat\'eriaux (LPEM), CNRS, Universit\'e Paris Sciences et Lettres.}
\author{B. Laburthe-Tolra$^1$}
\author{M. Robert-de-Saint-Vincent$^1$}
\email{martin.rdsv@univ-paris13.fr}

\affiliation{$^1$ Laboratoire de Physique des Lasers, CNRS, UMR 7538, Universit\'e Sorbonne Paris Nord,  F-93430 Villetaneuse, France}% \\
\affiliation{$^2$ Department of Physics and Engineering, Bloomsburg University, Bloomsburg, Pennsylvania }

%\pacs{Pacs Numbers}
%03.75.Hh	Static properties of condensates; thermodynamical, statistical, and structural properties
%03.75.Nt	Other Bose-Einstein condensation phenomena
%05.30.-d	Quantum statistical mechanics
%67.85.-d 	Ultracold gases, trapped gases
%03.75.Kk 	Dynamic properties of condensates; collective and hydrodynamic excitations, superfluid flow
%42.50.Tx	Optical angular momentum (quantum optics)
%47.37.+q 	Hydrodynamic aspects of superfluidity; quantum fluids
%67.85.De 	Dynamic properties of condensates; excitations, and superfluid flow
%37.10.Vz 	Mechanical effects of light on atoms, molecules, and ions
%37.10.De	Atom cooling methods
%37.10.Gh	Atom traps and guides
%37.10.Jk	Atoms in optical lattices
%67.85.Hj	Bose-Einstein condensates in optical potentials

\begin{abstract} % max 600 caractères
We introduce a spin-orbit coupling scheme, where a retro-reflected laser beam selectively diffracts two spin components in opposite directions. Spin sensitivity is provided by sweeping through a magnetic-field sensitive transition while dark states ensure that spontaneous emission remains low. The scheme is adiabatic and thus inherently robust. This tailored spin-orbit coupling allows simultaneous measurements of the spin and momentum distributions of a strontium degenerate Fermi gas, and thus opens the path to momentum-resolved spin correlation measurements on $SU(N)$ quantum magnets.
\end{abstract}

\date{\today}
\maketitle

Fermionic alkaline earth atoms are a new platform to study the physics of strongly correlated quantum many-body systems. An important interest arises from their large spin in their ground state ($F=9/2$ for strontium) combined with spin-independent interactions, that makes possible the study of quantum magnetism with an enlarged $SU(N)$ symmetry\,\cite{Gorshkov2010, Hermele2009, Wu2010, Cazalilla2014, Taie2010, Taie2012, Pagano2014, Hofrichter2016, Ozawa2018}. 
Interactions are spin-independent because the spin is entirely nuclear.
This comes with the additional consequence that these atoms are highly insensitive to magnetic fields. Thus, $SU(N)$ magnetism cannot be either controlled or probed using standard tools that make use of magnetic fields. Fortunately, these species possess narrow and ultra-narrow optical transitions, used in the context of optical atomic clocks \cite{Derevianko2011, Ludlow2015}, and which provide unique handles in the context of quantum magnetism\, \cite{Gorshkov2010, Cappellini2014, Zhang2014, Scazza2014}. One can use the tensor light shift associated with these narrow transitions to synthesize an effective magnetic field\,\cite{Dalibard2011, Mancini2015, Song2016, Lee2017} which can even be spatially tailored. This has been applied to separate the different spin components in a so-called Optical Stern-Gerlach (OSG)  \cite{Sleator1992,Taie2010,Stellmer2011}. 

Here, using the $^1S_0 \rightarrow$ $^3P_1$ transition of $^{87}$Sr atoms, we demonstrate a new scheme to separate the spin states of an $SU(N)$ degenerate Fermi gas. It is an adiabatic protocol in which the atoms in two well-defined Zeeman states are Raman-diffracted in well-defined directions using a simple retro-reflected beam, in $\sigma^+/\sigma^-$ configuration.
It inherently relies on spin-orbit coupling (SOC)\,\cite{Lin2009, Dalibard2011, Cheuk2012, Huang2016}, and benefits from the narrow linewidth of the transition\,\cite{Mancini2015,  Song2016, Lee2017}. %Livi2016,
Moreover, it involves nearly dark states, which further reduces spontaneous emission. Spin sensitivity is ensured by adiabatically sweeping through a magnetic-field dependent transition in the $^3P_1$ state $m_F$ manifold.

The scheme's basic principle evokes sawtooth-wave-adiabatic-passage cooling \cite{ Norcia2018, Bartolotta2018, Snigirev2019}, but is fully coherent and leads to acceleration.
For a gas with an initial momentum spread that is small compared to the recoil associated with the Raman process, 
two selected spin components can thus be spatially separated from the initial cloud after a time of flight. Therefore, their number and momentum distribution are then accessible by absorption imaging on the broad $^1S_0 \rightarrow$ $^1P_1$ transition, which dramatically increases the signal to noise ratio compared to direct imaging on $^1S_0 \rightarrow$ $^3P_1$\,\cite{Stellmer2011, Manai2020}. 
The full spin distribution can be measured in six successive experimental realizations, with minimal distortion of the momentum distributions of the selected spin states, in contrast to the OSG scheme.
We also demonstrate a generalized version of the scheme 
and realize a single-shot spin and momentum resolved picture of a five-component SU(5) Fermi gas. 
This sets up the possibility to probe correlations between spin and momentum in $SU(N)$ Fermi gases using correlation noise measurements\,\cite{Folling2005}, of interest e.g. to the study of quantum magnetism\,\cite{Bruun2009}.

\begin{figure}[h!]
\centering
  \includegraphics[width=\columnwidth]{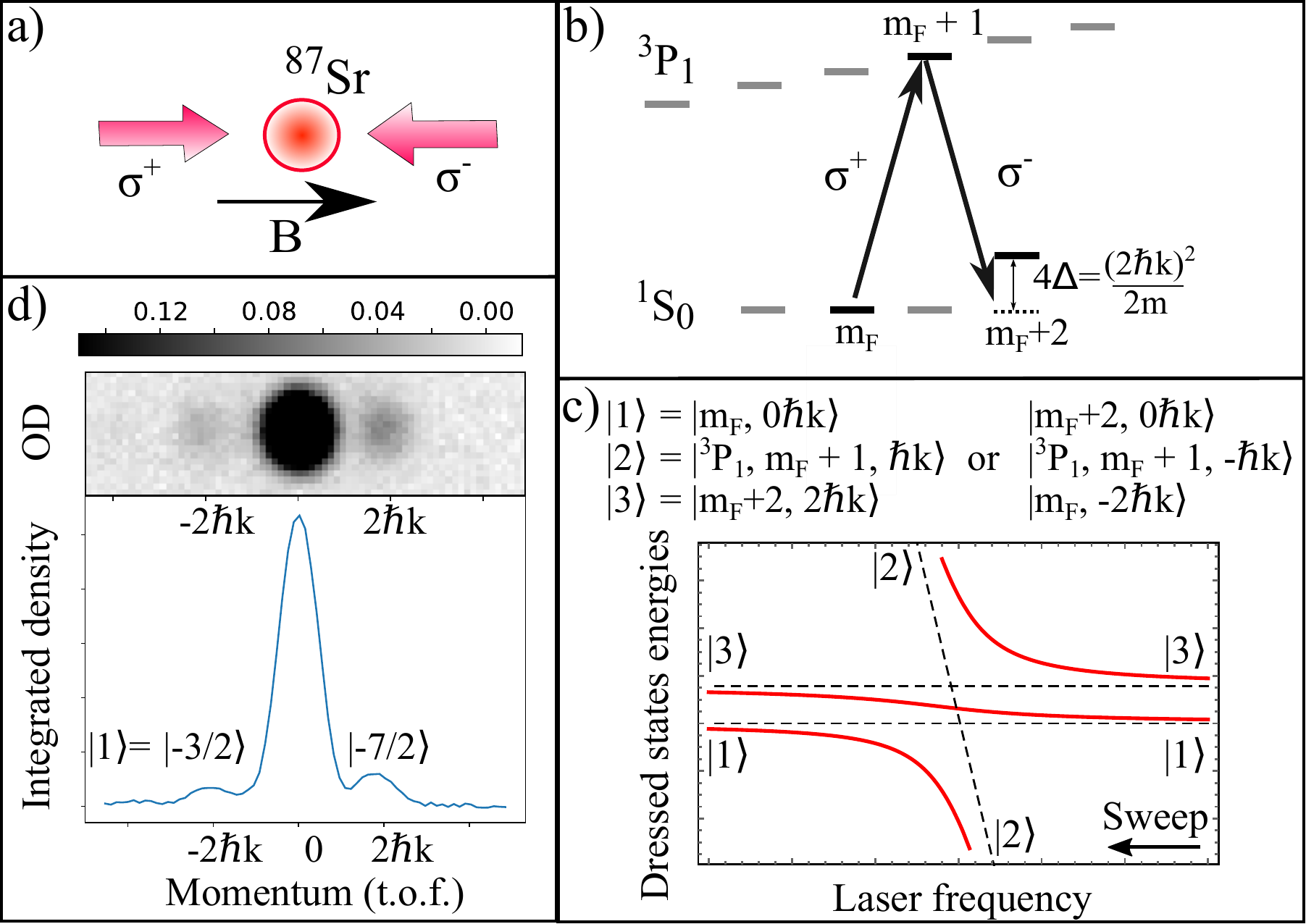}   
  \caption{Principle of the experiment. a) Geometry. The $^{87}$Sr cloud is exposed to a retro-reflected laser in $\sigma^+/\sigma^-$ configuration. %, is used to transfer two recoil momenta in opposite directions selectively to two spin states of $^{87}$Sr. 
b) Level scheme. A linear Zeeman effect in the excited state $^3P_1$ is used to selectively couple two spin states in the ground manifold, $|^1S_0, m_F\rangle$ and $|^1S_0, m_F+2\rangle$, quasi-resonantly via $|^3P_1, m_F+1\rangle$. The process, from either of the ground states, is associated with a momentum transfer $\pm 2 \hbar k$. 
  c) Dressed state energies in the rotating wave picture as a function of the laser frequency, with (solid lines) and without (dashed lines) the couplings with the light. A frequency sweep of the laser adiabatically connects one spin state at rest to the second spin state with recoil. 
    d)  Absorption image of the atomic cloud following the adiabatic momentum transfer and a time-of-flight, and corresponding 1D integrated profile. Two selected spin states, here $m_F = -3/2$ and $-7/2$ (initially), are extracted from the SU(10) Fermi sea and diffracted in opposite directions while the other eight spin states stay at rest. 
  }
 \label{fig:fig1}
 \end{figure}

The principle of our experiment is shown in Fig.~\ref{fig:fig1}. 
Atoms initially at rest in a selected spin state $m_F$ absorb one $\sigma^{+}$ polarized photon from a laser beam and re-emit one $\sigma^{-}$ polarized photon into the retroreflected beam. Denoting $\hbar k$ the single photon recoil momentum, this imparts a momentum $2 \hbar k$ to these atoms. 
%The process efficiently occurs when the state $|^1S_0, m_F\rangle$ is resonantly connected to $|^3P_1, m_F+1\rangle$. 
The process efficiently occurs when the states $|^1S_0, m_F\rangle$ and $|^1S_0, m_F+2\rangle$ are resonantly connected to $|^3P_1, m_F+1\rangle$. 
Therefore, simultaneously, the opposite process also occurs for atoms initially in $|^1S_0, m_F+2\rangle$, which are imparted $- 2 \hbar k$ along the beam axis.

We start by laying out a simple theoretical model describing the momentum transfer from a given $m_F$ state. The process relies on a three-level lambda system. We write its structure starting from the initial ground state $\left| 1 \right\rangle \equiv \left| ^1S_0, m_F; 0 \hbar k \right\rangle $ with spin state $m_F$ and momentum $p$ close to $0 \hbar k$, i.e. with $p \leq \hbar k_F < \hbar k$ where $\hbar k_F$ is the Fermi momentum. The intermediate state $\left| 2 \right\rangle \equiv \left|^3P_1,{m_F}+1; \hbar k \right\rangle $ is electronically excited with spin state $m_{F}+1$ and has gained one recoil momentum.  The final state $\left|3 \right\rangle \equiv   \left| ^1S_0,{m_F}+2; 2 \hbar k \right\rangle $ is in the ground state, with spin state $m_{F}+2$ and having gained $2 \hbar k $. 
In a dressed-atom picture the Hamiltonian describing the 3-level dynamics is given by  
\begin{equation}
	H = \hbar\begin{bmatrix}
0 				&	 \Omega_1/2  & 0 \\
\Omega_1/2 	&		\Delta-\delta & \Omega_2/2 \\
0 				&	 \Omega_2/2  & 4\Delta \\
\end{bmatrix}
\end{equation}
where $\Delta=\frac{\hbar k^2}{2m}$ is the recoil frequency and $\delta=\omega-\omega_0$ the frequency detuning between the bare level transition frequency $\omega_0$ and the laser frequency $\omega$. From $\left| 1' \right\rangle \equiv \left| ^1S_0, m_F+2; 0 \hbar k \right\rangle $, an independent three level structure can be identified, with the same Hamiltonian but for reversed $\Omega_1, \Omega_2$, and leading to $\left| 3' \right\rangle \equiv \left| ^1S_0, m_F; -2 \hbar k \right\rangle $.

In presence of the laser, the three states with respective energies $\hbar(0 ,\Delta -\delta,4 \Delta )$ give rise to three dressed states with energies shown on Fig.\,\ref{fig:fig1}c. 
One of those, $|\Psi_C (\delta) \rangle$, adiabatically connects $\left|1\right\rangle$ and $\left|3\right\rangle$ when the detuning $\delta$ is scanned through resonance. Furthermore, as the Rabi frequencies $(\Omega_1, \Omega_2) \sim \Omega$ increase,
the projection on the excited state $|\langle 2 |\Psi_C \rangle|^2$ decreases. A first-order perturbative expansion when $\Delta \ll \Omega$ and for $\delta=0$ shows that: 
\begin{equation}
\Psi_C(\delta=0) \approx \frac{1}{\sqrt{2}} \left(\left|1\right\rangle-\left|3\right\rangle\right)+2 \sqrt{2} \frac{\Delta}{\Omega} \left|2\right\rangle
\label{eqstate}
\end{equation}
which confirms that the state connecting $\left|1\right\rangle$ and $\left|3\right\rangle$  is nearly a dark state for  $\Delta \ll \Omega$. The main idea of our protocol is to follow adiabatically from  $\left|1\right\rangle$ to $\left|3\right\rangle$ while always remaining in this quasi-dark state, in order to reduce spontaneous emission. The adiabaticity of the process ensures its robustness. 

For $\delta \gg \Omega$, a first order perturbative expansion shows that $|\Psi_C(\delta)\rangle \simeq \left|1\right\rangle - \frac{\Omega^2}{16 \delta \Delta} \left| 3 \right\rangle $. Since our atoms are initially in $\left|1\right\rangle$, it is important for the success of the adiabatic transfer that this state is initially close to the dressed state $|\Psi_C(\delta)\rangle$, which requires:
\begin{equation}
\frac{\Omega^2}{16 \delta_{i} \Delta} \ll 1
\label{eqpropre}
\end{equation}
where $\delta_{i}$ is the initial value of $\delta$ when the light is switched on.    

Finally, adiabatic transfer from $\left|1\right\rangle$ to $\left|3\right\rangle$ is warranted if a Landau-Zener (LZ) criterion for adiabaticity is fulfilled, which qualitatively reads (see below for a refined analysis):
\begin{equation}
\frac{d \delta}{dt} \ll \Omega^2
\label{LZ}
\end{equation}
Equations (\ref{eqpropre}) and (\ref{LZ}) constitute the two main requirements for the success of adiabatic transfer. From those, and accounting for the excited state population (Eq. \ref{eqstate}), we find that the probability of spontaneous emission during a sweep scales as $\gamma \Delta / \Omega^2$. This scaling is confirmed by numerical simulations of the corresponding master equation, that accounts for spontaneous emission from $|2\rangle$. 

\begin{figure}[t!]
\centering
\includegraphics[width=\columnwidth]{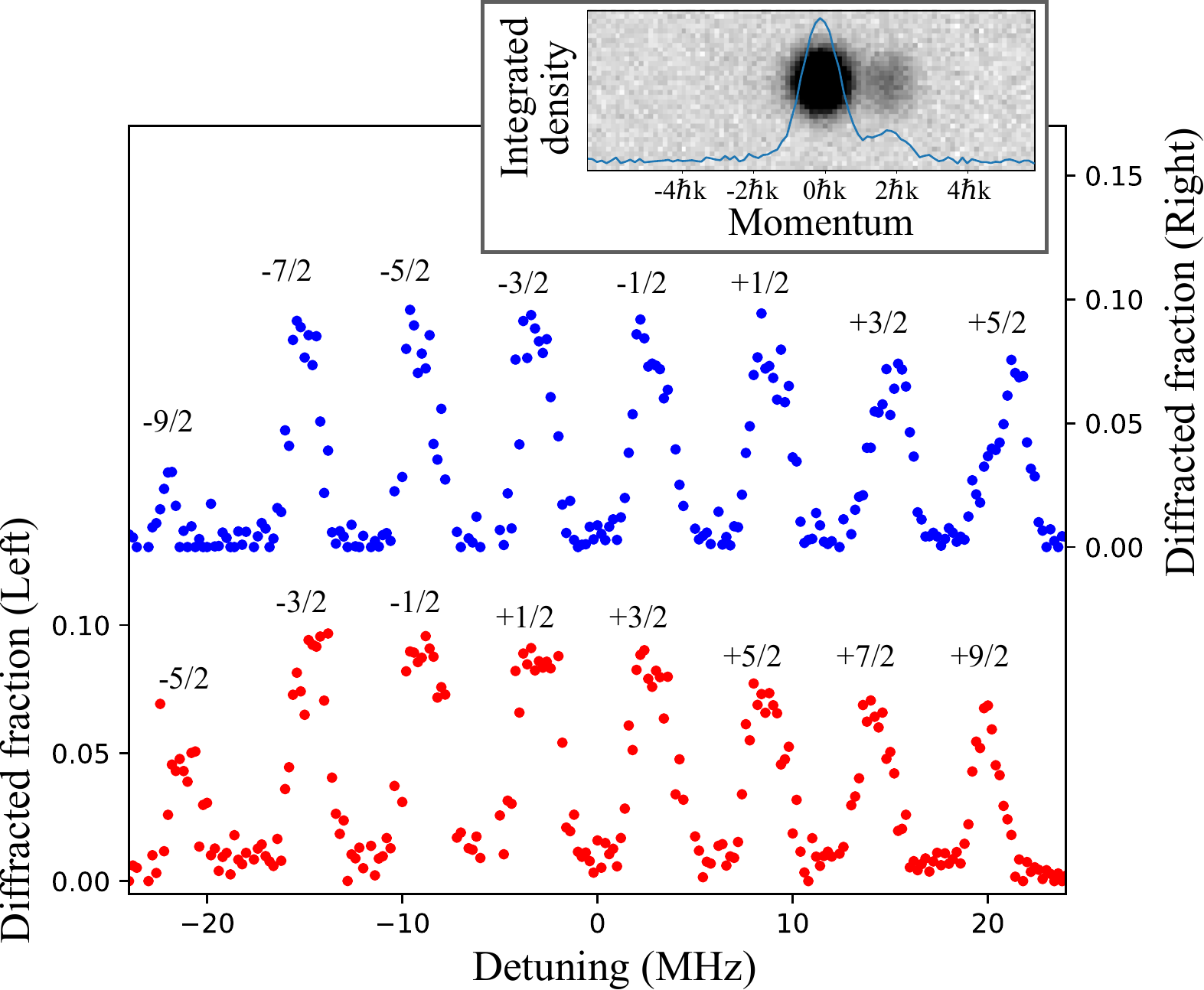}
    \caption{Fraction of the entire cloud diffracted in either direction, as a function of the frequency center of the chirp. At each diffraction peak, the two specific spin states selectively diffracted towards opposite momenta $\pm 2 \hbar k$ are labelled. 
Inset: momentum transfer at -3\,MHz from the spin states -3/2 and +1/2, where a prior $\sigma^-$ optical pumping sequence targeted +1/2 and -1/2 consecutively
(see Appendix C). The state +1/2, diffracted to $-2 \hbar k$, is therefore here empty.    
    }
 \label{fig:fig2}
 \end{figure}
 
We now turn to our experimental results. We prepare a 10-component spinor Fermi gas of about 25\,000 $^{87}$Sr atoms by forced evaporation in an optical dipole trap at a typical temperature $T\simeq 70\,$nK, with $T/T_F \simeq 0.3$ for each spin component, where $T_F$ is the Fermi temperature. Here, $k_B T_F \simeq 1.0 \times \hbar^2 k^2 / 2 m $, such that the velocity spread is below the recoil velocity $\hbar k/m$ at 689\,nm.
The apparatus and the production of degenerate Fermi gases is described in Appendix A. To implement the spin-selective momentum transfer, a vertical retro-reflected beam with $\sigma^+ / \sigma^-$ polarizations addresses the $|^1S_0, F= 9/2\rangle \rightarrow$ $|^3P_1, F=11/2\rangle$ transition at 689 nm, in the presence of a magnetic field of 16\,G along the beam propagation direction. 
We hold the atoms against gravity in a weakly confining (trapping frequency of 125\,Hz) horizontal dipole trapping beam during the momentum transfer.
The momentum distribution is then measured by absorption imaging on the $^1S_0\rightarrow$ $^1P_1$ transition after a 7-ms time-of-flight.

The 6.0 MHz linear Zeeman separation between adjacent magnetic states of $\left|^3P_1,F=11/2\right\rangle$ is much higher than the 7.4 kHz line-width of the transition. The laser intensity is 5 mW/cm$^2$, 
hence Rabi couplings are $2\pi \times 220\,$kHz, multiplied by the corresponding Clebsh-Gordan (CG) coefficients. The polarization is almost perfectly $\sigma^+$ in one direction (measured $>99\%$ in intensity) 
and $\sigma^-$ in the other. The laser frequency is ramped down, as sketched on Fig.\,\ref{fig:fig1}c. 
Each resonance corresponding to a given $|^3P_1, m_F+1\rangle$ state can therefore be used to selectively diffract two well-defined ground states $m_F, m_F+2$, as shown e.g. on Fig.\ref{fig:fig1}d, with almost perfect correlation between the initial spin state (here $m_F =-7/2$  and $-3/2$ ) and the direction in which it is diffracted. 

The adiabatic transfer is only performed as the laser is swept with \textit{decreasing} frequency. With the optimal parameters of sweep speed and span, we observe that it leads to little heating. The momentum distribution of the selected states is slightly narrower ($\sim 10\%$) along the diffraction axis, which we attribute to a reduced transfer efficiency for longitudinal momenta close to the recoil. The process also appears reversible: when combining a first sweep with decreasing frequency with a second sweep with increasing frequency, we observe that the cloud diffracted by the first sweep is almost entirely suppressed by the second one. In practice, the remaining atoms in the diffracted peak cannot then be distinguished from the background signal associated with the cloud of atoms that underwent spontaneous emission.
Finally, if the frequency is ramped up instead of down, we observe heating and a poor transfer efficiency, as atoms are adiabatically transferred to the excited state. Subsequently these atoms would return to the fundamental state by spontaneous emission, leading to a mean net momentum transfer of $\pm \hbar k$ and to heating.

In Fig.~\ref{fig:fig2} we present a complete spectrum of the diffracted atom numbers in both directions, as a function of the central frequency  of the laser sweep. Sweeps have a frequency span of 1.4\,MHz and a duration of 300\,$\mu$s. We count the number of diffracted atoms by fitting the time-of-flight pictures with three Gaussian distributions. We observe eight resonances, each corresponding to two different spin states being diffracted in opposite directions. Each of these can thus be used to selectively measure the atom number and momentum distribution in two spin states. According to numerical simulations for our parameters, the efficiency is approximately the same on all lines ($\sim 80\%$ diffracted, and 10 to 20\,$\%$ undergoing spontaneous emission - typically five times less than if the evolution were incoherent\,\cite{Noh2016}), but for the two extreme lines that should be less efficient ($\sim 50\%$ diffracted) due to pathologically small CG coefficients. This is in good agreement with the fact that roughly 75$\%$ of the total atom number is accounted for by the various diffraction resonances. 

Given the poorer efficiency of the two outlying lines at $\pm$21\,MHz, they are not as reliable as the others for an atom number calibration. The extreme asymmetry, by a factor 6, between the two CG coefficients of each of these lines is also responsible for a MHz shift between the diffraction resonances in opposing directions. Nevertheless, from the data it seems that the state $m_F = -9/2$ is weakly populated. On all the other lines, our scheme is a sensitive tool to measure the populations of the various spin states, for example to calibrate the preparation of ad hoc spin distributions by optical pumping using the intercombination line\,\cite{Stellmer2011}, as shown e.g. in the inset of Fig.\,\ref{fig:fig2}.

The typical transfer efficiency could be brought around e.g. $95\%$ by using a four times larger intensity, thus reducing spontaneous emission from $|\Psi_C\rangle$. This would also require to increase by at least the same factor the chirp span, as well as the magnetic field to keep the Zeeman splitting in $^3P_1$ larger than the chirp span. However, for such fields ($\sim 60\,$G), the linear Zeeman effect in the ground state exceeds the recoil energy.
This can be easily compensated for, in our vertical implementation of the scheme, by a small free-fall time ($g_I \mu_0 B / \hbar k g \sim 1\,$ms, with $g$ the gravitational acceleration, 
$g_I \simeq 1.09 \mu_N/\mu_0 I$ the Lande factor of $^{87}$Sr in its ground state, $I=9/2$ its nuclear spin, and $\mu_N$ the nuclear magneton \cite{Olschewski1972}).

\begin{figure}[t!]
\centering
  \includegraphics[width=\columnwidth]{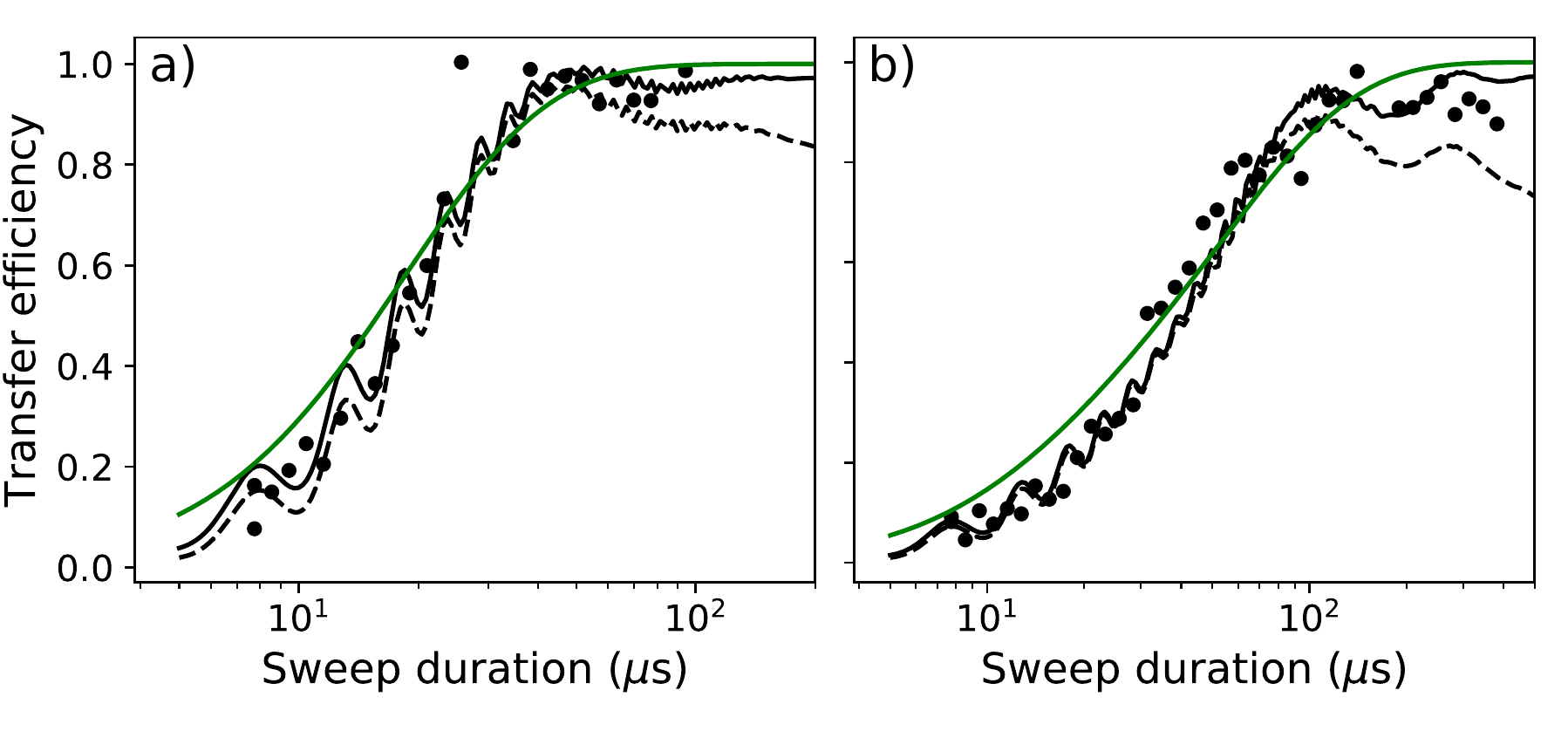}
  \caption{Adiabatic transfer dynamics, a) from $|-3/2, 0 \hbar k \rangle$ to $|+1/2, 2 \hbar k \rangle$ at -3 MHz, and b) from $|-7/2, 0 \hbar k \rangle$ to $|-3/2, 2 \hbar k \rangle$ at -15 MHz.
The points are measured transfer efficiencies as a function of ramp duration, assuming a 10$\%$ occupation of the spin states. The black lines result from numerical simulations without adjustable parameter: the black solid line accounts for the atoms collecting two recoils or having undergone spontaneous emission, and the dashed line is the fraction that underwent precisely two recoils. The data are finally compared to the LZ scaling $P_{adiab}^{LZ}(\Omega_1) \times P_{adiab}^{LZ}(\Omega_2)$ (solid green line).
}
\label{fig:fig3}
 \end{figure}

We now characterize the dynamics of the momentum transfer, and show that a Landau-Zener (LZ) scaling provides a good description of the optimal timescales.
Fig.\,\ref{fig:fig3}, as an example, shows the diffraction efficiencies for the line at -3\,MHz, coupling the spin state -3/2 towards +1/2 (panel a), and at -15 MHz, coupling the spin state -7/2 towards -3/2 (panel b), as a function of the sweep duration.
Both data correspond very well to our numerical model, solving the master equation including spontaneous emission. Here we assume a 10$\%$ population in each state, motivated by our observations of Fig.\,\ref{fig:fig2}; otherwise there is no adjustable parameter. 
Residual oscillations in the numerical model show that the condition in Eq.\,\ref{eqpropre} is not perfectly respected. 
Unlike, e.g., \cite{Olson2014}, the LZ problem in our SOC mechanism remains inherently three-level, and the exact adiabaticity timescales depend on the relative values of $\Omega_1, \Omega_2, \Delta$\, \cite{Carroll1986, Yurovsky1999, Band2019}.   
We first introduce the usual two-level LZ transfer probability:
\begin{equation}
P_{adiab}^{LZ}(\Omega) \simeq 1 - \exp(-2 \pi \frac{\Omega^2}{4 \dot \delta})
\label{eq:PLZ}
\end{equation}
where $\hbar \Omega$ is the gap in a two-level avoided crossing and $\dot \delta$ is the chirp rate. 
In both data presented in Fig.\,\ref{fig:fig3}, we find that the sweep-duration dependence follows $P_{adiab}^{LZ}(\Omega_1) \times P_{adiab}^{LZ}(\Omega_2)$ \cite{Carroll1986}. The smallest Rabi coupling is thus the limiting quantity for the adiabatic timescale. %, which is especially striking on the data of Fig.\,\ref{fig:fig3}b, where $\Omega_1 \sim \Omega_2/3$. 
In both data sets, the smallest coupling is in the first leg of the diffraction process, i.e. the absorption process. The non-diffracted atoms are then mostly left at rest. For the opposite selected spin state, in particular when the CG asymmetry is extreme (large $|m_F|$), sweeps faster than the adiabaticity criterion can lead to a large level of spontaneous emission.

Let us now discuss the experimental limits to spin selectivity. Diffraction of the wrong spin states can arise either from imperfect circular polarization of the laser beams, or from non-collinear propagation of the lasers and external magnetic field. Both of these correspond to  deviations from the $\sigma^+ / \sigma^-$ polarization along the magnetic field axis, that may enable $\pm 2 \hbar k$ momentum transfer from unintended initial $m_F$ states, and thus break the perfect correlation between the initial spin state and the recoil direction. 
We estimate that a light circularly polarized to better than $99\%$ in intensity and an angle between the magnetic field and beam axes $\theta <6^{\circ}$ suppresses the diffraction efficiencies on unintended spin states down to a few percent, provided the sweep timescale is kept weakly above the adiabatic timescale for the desired transition (see Appendix B).
To confirm this, we selectively emptied  one or two spin states using optical pumping\,\cite{Stellmer2011} (see Appendix C), prior to evaporation and adiabatic momentum transfer - see e.g. the inset of Fig.\,\ref{fig:fig2}. Any measured diffracted cloud must be attributed either to imperfect optical pumping or to imperfect selectivity of the transfer scheme. The weak measured signals, $\sim 1\,\%$ of the total atom number, provide a worst-case scenario estimate for the diffraction efficiency from the wrong states $\sim 10\%$.

%%%%%%%%%

We now compare our two-photon scheme to more traditional ones. As compared to an off-resonant Raman $\pi$-pulse, our scheme offers the robustness to power-drifts inherent to adiabatic schemes. 
It is similar to a stimulated Raman adiabatic passage (STIRAP) procedure \cite{Vitanov2017}. However, 
due to the very small frequency difference between the two legs of the $\Lambda$ scheme, STIRAP would not be immune to imperfect polarizations, that would result in finite spontaneous emission. 
Furthermore, while STIRAP can be tuned to efficiently diffract one state, e.g.  $|m_F, 0\hbar k\rangle$ towards $|m_F+2, 2\hbar k\rangle$, during this pulse the state $|m_F+2, 0\hbar k\rangle$ will undergo a coupling sequence resulting mostly in spontaneous emission.
Our scheme is attractive in its simplicity, robustness, and its roughly symmetric efficiency for two spin states at once - that we will now use for a generalization of the scheme.

\begin{figure}[t!]
\centering
  \includegraphics[width=\columnwidth]{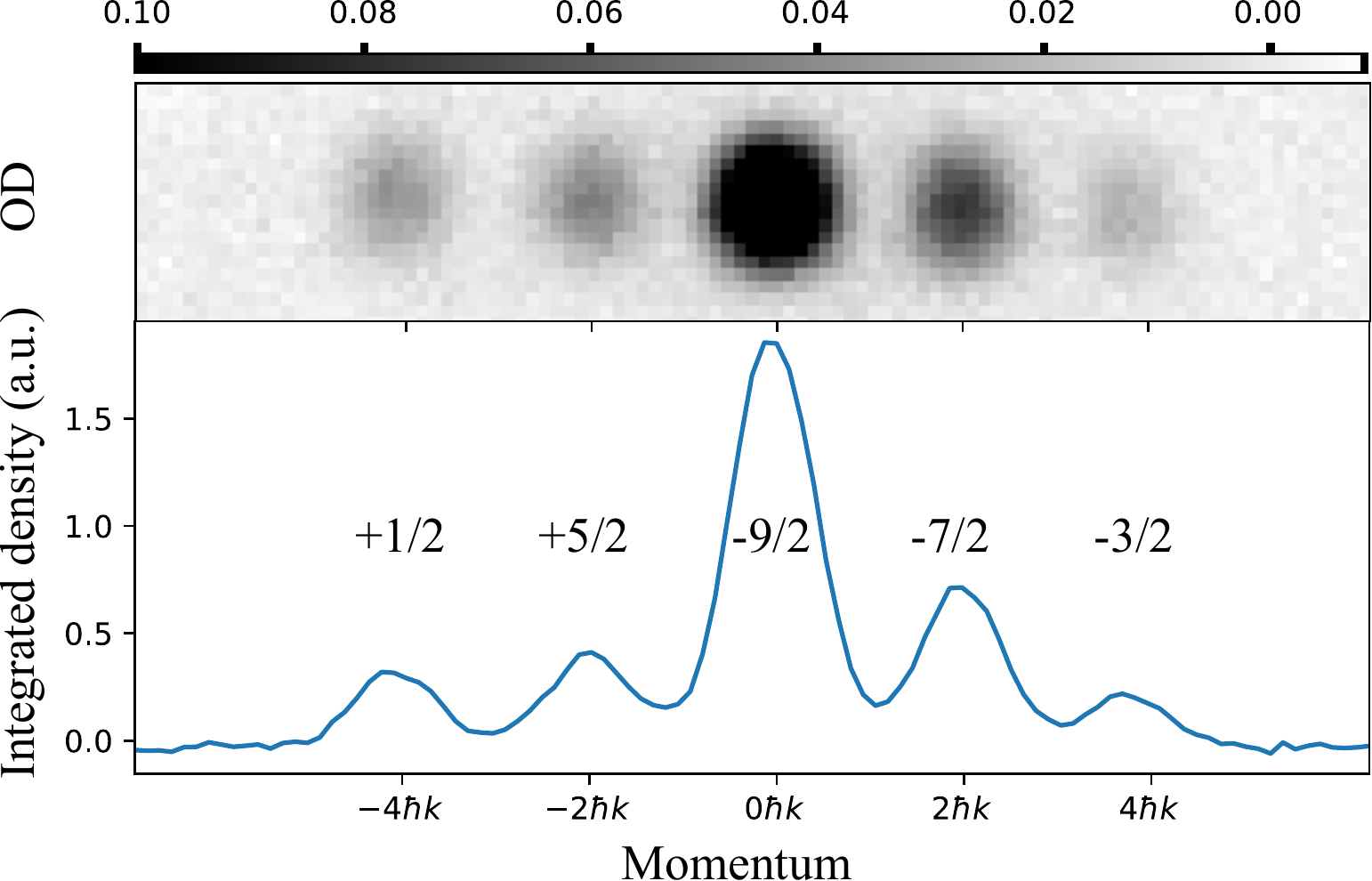} %2020_02_28
  \caption{Three-pulse measurement of the five spin states in a SU(5) Fermi gas prepared by optical pumping, providing a complete single-shot spin-resolved momentum distribution. Each cloud is assigned to its initial spin state. The central cloud, that in principle should be mostly $m_F=-9/2$, is admixed with residuals from the other states given our moderate transfer efficiency. 
  }
 \label{fig:3pulse}
 \end{figure}

In this last experiment, we demonstrate selective momentum transfer to four spin states in a single realization - here $\{-7/2, -3/2, +1/2, +5/2\}$, that acquire four different momenta. This involves a three pulse sequence. To prevent a detrimental effect of the trapping potential, with period 8\,ms, we reduce the sweep durations by a factor of up to two. The need for a trapping potential could be prevented by realizing the momentum transfer horizontally. The three pulses sweep consecutively the lines that connect i) $-3/2$ and $+1/2$, ii) $+1/2$ and $+5/2$, and iii) $-7/2$ and $-3/2$.
The atoms initially in the states $-7/2$ and $+5/2$ are sensitive, respectively, to the third and second pulse, and diffracted to $+2 \hbar k$ and $-2 \hbar k$, respectively. The atoms initially in the states $-3/2$ and $+1/2$ each react to two of the pulses and are diffracted to $+ 4 \hbar k$ and $- 4 \hbar k$ respectively. For example, the first pulse transfers the atoms in $|-3/2, 0 \hbar k \rangle$ to $|+1/2, 2 \hbar k \rangle$; the second pulse then acts on these same atoms and transfers them to $|+5/2, 4 \hbar k \rangle$.
We present the result of this procedure in Fig.\,\ref{fig:3pulse} for a cloud initially prepared in a mixture of the 5 states $\{-9/2, -7/2, -3/2, +1/2, +5/2 \}$ by optical pumping (see Appendix C).
Each of the five momentum-separated clouds is assigned a spin state. Given our moderate (but improvable) transfer efficiency, the central cloud remains significantly ($\sim 50\%$ in principle) contaminated by the other spin states. Nevertheless, 
%despite the poor but improvable central cloud spin purity, 
this picture demonstrates a single shot, spin and momentum resolved measurement of a degenerate SU(5) Fermi gas, currently inaccessible to established techniques.

Our scheme is an optimal light-engineered SOC scheme in that  
the momentum kick, set by the wave-length, is the highest possible. 
As a consequence, the measurement is fast, and can be accomplished with very large laser beams (1 cm radius in practice). Thus, distortion of the cloud is minimal, so that each image gives access to both the momentum and the spin distributions, opening the path to $e.g$ correlation measurements \cite{Bruun2009}. 
Finally, we also point out that our scheme provides a swap between states $m_F$ and $m_F+2$, which is one of the main one-body quantum gates corresponding to the generators of the $SU(N)$ group needed to fully characterize $SU(N)$ quantum magnets \cite{Merkel2008, Sosa-Martinez2017, Leroux2018}. 

\begin{acknowledgments}
We acknowledge B. Pasquiou and F. Schreck for fruitfull discussions helping in the design of the experimental apparatus, and  L. Vernac and P. Pedri for critical reading of the manuscript.
This research was funded by the Agence Nationale de la Recherche (projects ANR-18-CE47-0004 and ANR-16-TERC-0015-01), the Conseil R\'egional d'Ile-de-France, Institut FRancilien des Atomes Froids, DIM Nano'K (projects METROSPIN and ACOST), DIM Sirteq (projects SureSpin and Suprisa), and Labex FIRST-TF (project CUSAS). 
\end{acknowledgments}

\appendix
\section{Production of degenerate $^{87}$Sr Fermi gases}
\label{sec:setup}

We here describe our experimental apparatus and the production procedure of degenerate $^{87}$Sr Fermi gases, which share similarities with other experiments \cite{DeSalvo2010,StellmerBEC, KillianBEC,Stellmer2013}. 

Our oven is kept at 790\,K, and releases an effusive beam of all Sr isotopes. The fermionic isotope $^{87}$Sr natural abundance is 7\%. The effusive beam is transversely-cooled by 2D molasses on the $^1S_0 \rightarrow$ $^1P_1$ transition (461 nm, $\Gamma_{blue} / 2\pi = 30$ MHz), then decelerated in a 80 cm Zeeman slower with $550$\,m/s capture velocity. After that first cooling stage, $^{87}$Sr atoms are captured in a 3D MOT, still using the $^1S_0 \rightarrow$ $^1P_1$ transition. Its typical temperature is around 1 mK, and it has a life time of a few tens of ms due to optical pumping to the long-lived metastable $^3P_2$ state via the $^1D_2$ state\,\cite{Kurosu1992}. Because of this, laser-cooled atoms in positive Zeeman states accumulate in the $^3P_2$ state and remain magnetically trapped. We use a 51 G/cm gradient during this MOT phase, which is three to four times higher than gradients usually used in alkali MOTs, in order to reduce the volume of mK-temperature magnetically trapped cloud.

After 5 seconds of accumulation, we repump $20\times10^6$ atoms in the fundamental $^1S_0$ state using a 50 ms laser pulse, resonant with the $^3P_2  \rightarrow$ $^3D_2$ transition at 403 nm. This laser is rapidly  modulated  ($\approx$5\,ms) over roughly 2\,GHz, in order to repump at least the F = 13/2 and F = 11/2 hyperfine states of $^3P_2$ into $^3P_1$, from which they decay to $^1S_0$.
Simultaneously,  the 51 G/cm gradient is abruptly brought down to 1.08 G/cm, % timescale 10 ou 20 ms?
a blue MOT is applied at a very strongly reduced power to help the spatial recompression of the repumped cloud, and we turn on a second MOT that uses the narrow intercombination line $^1S_0, F=9/2\rightarrow$ $^3P_1, F=11/2$\,\cite{Katori1999, Mukaiyama2003} (689 nm, $\Gamma_{red} / 2\pi = $ 7.4 kHz). After about 20 ms the blue MOT is then fully turned off, and closely after that the repumper light is also switched off, such that only the 689\,nm narrow-line MOT light is maintained.
This MOT is unstable because the Lande factor $g_e$ in the excited state is much larger than the Lande factor in the ground state $g_g$ : $g_e/g_g \gg F/F-1$. Therefore, as developed in ref.\,\cite{Mukaiyama2003, DeSalvo2010}, we realize a dynamical MOT that combines a ``trapping laser'' on $^1S_0, F=9/2\rightarrow$ $^3P_1, F'=11/2$ and a  ``stirring'' laser on $^1S_0, F=9/2\rightarrow$ $^3P_1, F'=9/2$, whose main role is to improve the randomization of magnetic-state populations.
We point-out that while this difficulty is common to any MOT with $g_e \gg g_g$ (including the broad-band MOT in Sr on the $^1S_0 \rightarrow$ $^1P_1$ transition), the difficulty is extreme here because of the narrow linewidth of the $^1S_0 \rightarrow$ $^3P_1$ transition. Because of this, atoms can be trapped for long times into states where they do not absorb light, such that randomization of the Zeeman population is crucial to allow efficient cooling.

Similarly to \cite{Katori1999, Mukaiyama2003}, the narrow line cooling stage is performed in four steps. Initially, the two 689 nm lasers are turned on with enough intensity $I_{laser}$ to largely saturate the transition: $I_{laser} \simeq 1000\times I_{sat}$, where $I_{sat}=3$ $\mu$W/cm$^2$. The trapping and stirring lasers are also spectrally-broadened via a triangle-shaped frequency modulation, respectively at 35\,kHz with 4.24 MHz full width  and at 24\,kHz with 6.2 MHz full width, to increase the velocity and spatial capture range of the trap. The average detuning for the trapping laser during that stage is 2.25 MHz, while that of the stirring laser is 3.2 MHz. This first step of spectrally-broadened cooling lasts for 240\,ms.

In a second step, over the course of 100 ms, the modulation width of the trapping laser is linearly brought down to 240 kHz, while the average red detuning decreases to 252 kHz. The modulation width of the stirring laser is brought down to 1 MHz, with an average red detuning of 628 kHz. 
This reduces the size of the cloud to allow for better mode matching with the optical dipole trap that is turned on later in the sequence.

In a third step, both lasers' intensities are decreased down to a few $I_{sat}$ over 150 ms, while increasing the magnetic gradient from 1.08 G/cm to 2.1 G/cm, once again to compress the cloud in preparation for the optical dipole trap loading. Simultaneously, both laser frequencies are ramped to higher frequencies by 180\,kHz.
While finishing this manuscript, we realized that this ramp brought part of the modulated frequency spectrum blue-detuned. This indicates that this stage of our sequence could be better optimized. Nevertheless, this produces a sample of $\sim 10^7$ atoms at a temperature $\sim 10\,\mu$K.

\begin{figure}[t!]
\centering
\includegraphics[width= \columnwidth]{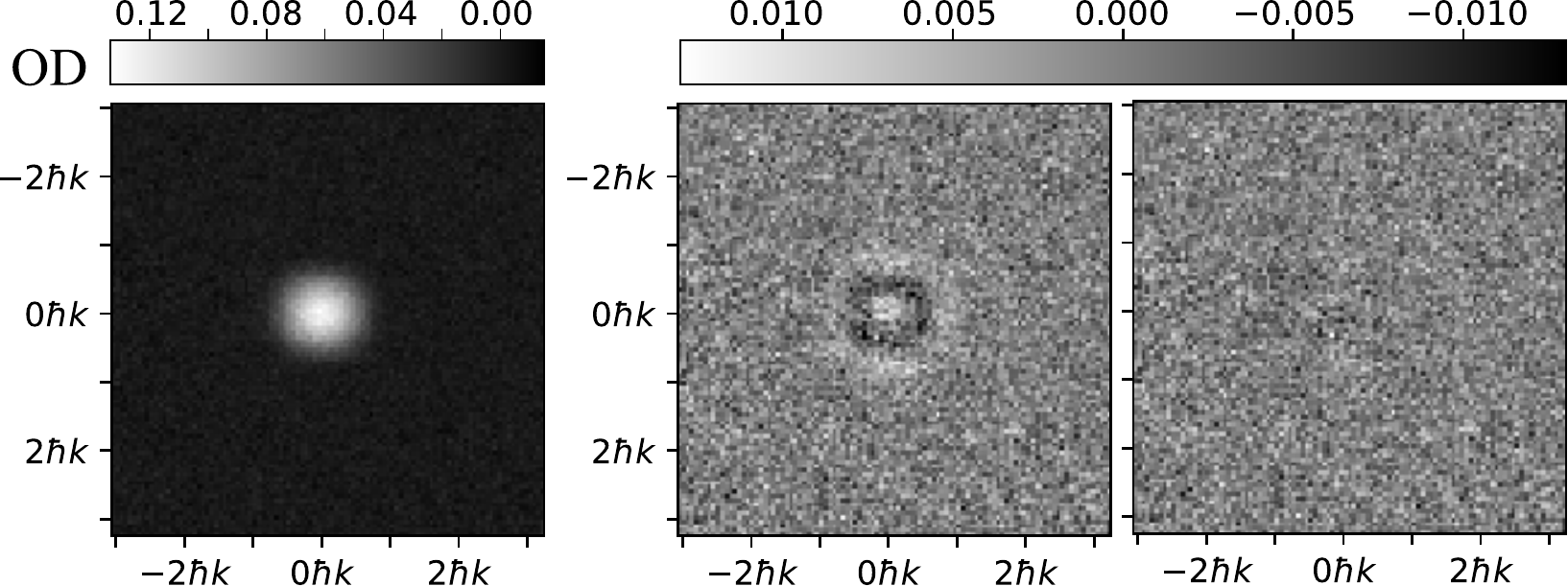}
\caption{Momentum distribution of a degenerate SU(10) Fermi gas, with $10^4$ atoms. Left: Time of flight picture. 
Right: 2D fit residuals, from left to right, from a classical Boltzmann distribution and from a non-interacting Fermi distribution \cite{Ketterle2008}. From the latter fit we estimate $T = 31$\,nK and $T/T_F =0.2$.}
\label{fig:Fermi}
 \end{figure}

Finally, the lasers are set to a single frequency, with a red detuning of 184 kHz for the trapping laser, and 36 kHz for the stirring laser. We simultaneously turn on a crossed optical dipole trap (ODT), using a 1070 nm ytterbium-doped, 20 W fiber laser, that combines two beams. The first beam, 10\,W,  is about $1^\circ$ away from horizontality, 
such that atoms evaporating from the crossing region into the horizontal beam can be extracted efficiently.
This beam is elliptical, with waists 160$\pm 5$\,$\mu$m horizontally and 45$\pm 5$\,$\mu$m vertically on the atoms. The second beam, 5\,W, is at 30$^\circ$ from the vertical, with a waist of 80$\pm 10$\,$\mu$m on the atoms.
Both beams differ in frequency by a few MHz to prevent interference. They have linear orthogonal polarizations, which minimizes not only interference but also the spin-dependence of the Stark shift on the cooling transition $^1S_0 \rightarrow$ $^3P_1$. The ODT creates a dark spot effect, although not as strong as in e.g.\,\cite{Stellmer2013}, such that trapped atoms perceive the cooling light as farther red-detuned by several $10^5$\,Hz (spin-dependent). The intensity of both cooling lasers are then further decreased using a linear ramp of 75 ms, while their frequencies are increased by 270\,kHz such that laser cooling becomes more efficient in the dipole trap potential - although at the cost of becoming blue-detuned for atoms away from the dipole trap.

The two 689 nm lasers are then switched off. The ODT, loaded with up to $10^6$ atoms at $4\,\mu$K mostly in the crossing region, is kept at a constant depth for 130 ms, then ramped down over 5\,s to produce a degenerate Fermi gas by forced evaporation. At the end of this procedure, the trap depth is determined by the gravitational sag, such that the horizontal beam intensity controls the temperature while both beams contribute to the trap confinement frequencies ($\sim 150$\,Hz). %and, mostly, the vertical confinement frequency ($\sim 125$\,Hz), while the vertical beam mostly controls the two other confinement frequencies ($\sim 170$\,Hz). 
We prepare clouds down to 30 nK, at $\frac{T}{T_F}\simeq 0.2$ as extracted from the shape of the momentum distribution (see Fig.\,\ref{fig:Fermi}), with about $10^4$ atoms. The experiments described in the main text are with a warmer cloud ($2.5\times10^4$ atoms at $\frac{T}{T_F}\simeq 0.3$).

 \section{Influence of polarization imperfections on the spin selectivity}
\label{sec:appendixpolar}

To account for the effect of imperfect polarization of our lasers on the spin selectivity, we consider the
three possible polarizations for each beam:  $\left( (\sigma_L^+, \sigma_L^-, \pi_L)\otimes(\sigma_R^+, \sigma_R^-, \pi_R)\right)$, where we annotate L (Left) and R (Right) the two beams. There are thus nine possible combinations of polarizations that could be involved in a two-photon transition with recoil.
The laser beams are mostly polarized $( \sigma_L^+\otimes \sigma_R^-)$, with Rabi couplings $(\Omega_L, \Omega_R) \simeq \Omega$ along this proper polarization.
We consider small deviations from these circular polarizations such that the Rabi frequencies in $\sigma_L^-$ or $\sigma_R^+$ scale as $\epsilon \Omega$ with $\epsilon \ll 1$. Likewise, we consider small angles $\theta$ between the magnetic field axis and the laser propagation axis, such that the Rabi frequencies in $\pi_L$ or $\pi_R$ scale as $\theta \Omega$.

We take the generic example where the spin states $m_F$, $m_F+1$ and $m_F+2$ in the ground state are coupled to one state $m_F+1$ in $^3P_1$.   The second column of the table shows how the different processes scale as a function of $\epsilon$ and $\theta$. Each box in the table contains the final spin state and the order in which it interacted with the two beams (absorption $\rightarrow$ stimulated emission). We consider the case where one wants to measure $m_F$ atoms, that should be diffracted by the $( \sigma_L^+\otimes \sigma_R^-)$ component in the left-to-right direction ($L\rightarrow R$). We must therefore make sure that all the processes diffracting atoms from the wrong initial state ($m_F+1$ or $m_F+2$) in the $L\rightarrow R$ direction are fully negligible. At first orders in $(\epsilon, \theta)$, there are two dominant processes, shown in the fourth and sixth line of the table, corresponding to $m_F+2$ atoms interacting with $\left( \sigma_L^-\otimes \sigma_R^-\right)$, and $m_F+1$ atoms interacting with $\left( \pi_L \otimes \sigma_R^-\right)$ light.
If the intended adiabatic process is realized with a chirp rate satisfying by only a moderate margin the LZ criterion, e.g.  $P^{LZ}_{adiab} \simeq 1- \exp{\left(- 2\pi \Omega^2/4\dot\delta\right)} \simeq 0.95$, keeping $\epsilon <0.1$ and $\theta <0.1$ ensures for the two dominant unintended processes a near-total failure of adiabaticity from the absorption process, with $\sim 97 \%$ of the atoms left at rest. These two conditions correspond to a light circularly polarized to better than 99$\,\%$ in intensity, and an angle $\theta < 6^{\circ}$.

\begin{widetext}
\begin{tabular*}{2 \columnwidth}{|c|c|c|c|c|}
   \cline{1-5}
   Polarization & Scaling & Initial state: $m_F$ & Initial state: $m_F+1$ & Initial state: $m_F +2$  \\
   \cline{1-5}
     $( \sigma_L^+\otimes \sigma_R^+)$ & $\epsilon$ & $m_F$ $L\rightarrow R$ ; $m_F$ $R\rightarrow L$  & $\oslash$  & $\oslash$ \\
   \cline{1-5}
     $( \sigma_L^+\otimes \sigma_R^-)$ & 1 & $m_F+2$ $L\rightarrow R$ & $\oslash$ & $m_F$ $R\rightarrow L$\\
   \cline{1-5}
	   $( \sigma_L^-\otimes \sigma_R^+)$ & $\epsilon^2$ & $m_F+2$ $R\rightarrow L$ & $\oslash$ & $m_F$ $L\rightarrow R$ \\
   \cline{1-5}
	   $( \sigma_L^-\otimes \sigma_R^-)$ & $\epsilon$ & $\oslash$ & $\oslash$ & $m_F+2$ $L\rightarrow R$ ; $m_F+2$ $R\rightarrow L$\\
   \cline{1-5}
	   $( \pi_L \otimes \sigma_R^+)$ & $\theta \epsilon$ & $m_F+1$ $R\rightarrow L$ & $m_F$ $L\rightarrow R$ & $\oslash$ \\
   \cline{1-5}
	   $( \pi_L \otimes \sigma_R^-)$ & $\theta$ & $\oslash$ & $m_F+2$ $L\rightarrow R$ & $m_F+1$ $R\rightarrow L$\\
   \cline{1-5}
	   $( \sigma_L^+\otimes \pi_R)$ & $\theta$ & $m_F+1$ $L\rightarrow R$ & $m_F$ $R\rightarrow L$ & $\oslash$ \\
   \cline{1-5}
	   $( \sigma_L^-\otimes \pi_R)$ & $\theta \epsilon$ & $\oslash$ & $m_F+2$ $R\rightarrow L$ & $m_F+1$ $L\rightarrow R$ \\
   \cline{1-5}
	   $( \pi_L\otimes \pi_R)$ & $\theta^2$& $\oslash$ & $m_F+1$ $L\rightarrow R$ ; $m_F+1$ $R\rightarrow L$  & $\oslash$ \\
   \cline{1-5}

\end{tabular*}
\end{widetext}

\section{Optical pumping and preparation of a 5-component gas.}
\label{sec:appendixpumping}
Optical pumping is realized on the intercombination line $|^1S_0, F=9/2\rangle \rightarrow$ $|^1S_0, F=11/2\rangle$, with an intensity  $\sim\,30 I_{sat}$, prior to the evaporative cooling. 
The laser is $\sigma^-$ polarized along a 6.8\,G magnetic field, such that the lines from each Zeeman sublevel $|^1S_0, m_F \rangle \rightarrow$ $|^3P_1, F=11/2, m_F-1\rangle$, are separated by 2.55 MHz and well resolved. Similarly to \cite{Stellmer2011}, we can then selectively pump atoms out of an arbitrary set of spin states.
The picture in the inset of Figure 2 shows that this procedure is efficient in removing atoms from unwanted spin states (while increasing the population in the other states), such that an arbitrary spin population can be designed.

To prepare a five-component Fermi gas, we realize a sequence of five $\sigma^-$ optical pulses, emptying, consecutively and in the following order, the states $m_F = +9/2, +7/2, +3/2, -1/2, -5/2$. To minimize the sensitivity to frequency or magnetic field drifts, each pulse, lasting 5\,ms, is swept in frequency over 100\,kHz through the resonance.

%\bibliography{References}
%\bibstyle{prsty}


\end{document}